# The Design and Operation of Digital Platform under Sociotechnical Folk Theories

*Emergent Research Forum (ERF) Paper*

**Jordan W. Suchow**
Stevens Institute of Technology
jws@stevens.edu

**Lea Burton**
Stevens Institute of Technology
lburton@stevens.edu

**Vahid Ashrafimoghari**
Stevens Institute of Technology
vashraf1@stevens.edu

## Abstract

We consider the problem of how a platform designer, owner, or operator can improve the design and operation of a digital platform by leveraging a computational cognitive model that represents users' folk theories about the platform as a sociotechnical system. We do so in the context of Reddit, a social-media platform whose owners and administrators make extensive use of shadowbanning, a non-transparent content moderation mechanism that filters a user's posts and comments so that they cannot be seen by fellow community members or the public. After demonstrating that the design and operation of Reddit have led to an abundance of spurious first-party suspicions of shadowbanning in cases where the mechanism was not in fact invoked, we develop a computational cognitive model of users' folk theories about the antecedents and consequences of shadowbanning that predicts when users will attribute their on-platform observations to a shadowban. The model is then used to evaluate the capacity of interventions available to a platform designer, owner, and operator to reduce the incidence of these false suspicions. We conclude by considering the implications of this approach for the design and operation of digital platforms at large.

## Introduction

A digital platform is a website or mobile application that enables its users to interact with each other or with platform operators (De Reuver, Sørensen, & Basole, 2018). The choices that designers, owners, and operators make with respect to the design and operation of a digital platform affect the user's acceptance and usage of the platform (Venkatesh, Morris, Davis & Davis, 2003). In some cases, these choices may lead the user to develop a folk theory about how the platform works (DeVito et al., 2018). A folk theory is a layperson's understanding of how something works. It is typically an oversimplification of reality, but it can nonetheless be useful for guiding behavior. For example, many people have a folk theory about how a disease is transmitted (Motta & Callaghan, 2020). This folk theory may be oversimplified or even incorrect, but it can nonetheless be useful for guiding hygiene behavior. In the context of digital platforms, folk theories about how the platform works can guide the user's behavior on the platform. For example, a user may develop a folk theory that the platform is designed to promote certain content over other content. This folk theory may lead the user to believe that the platform is biased against certain groups of people. If the user identifies with one of those groups, the user may avoid posting certain types of content on the platform, leading to an indirect chilling of speech that flows from the user's mental model of the platform's operation.

When designers, owners, or operators of a digital platform aim to correct users' misperceptions about a platform's operation, they do not have direct access to the users' folk theories about the platform. Rather, they must rely on indirect evidence, such as user behavior on the platform or in certain cases the results of surveys or other instruments aimed at understanding the userbase. In some cases, this indirect evidence may be sufficient for the platform designer, owner, or operator to identify a user's folk theory and take steps to correct it. In other cases, however, the indirect evidence may be ambiguous, making it difficult to identify the user's folk theory with certainty. In these cases, a computational cognitive model of the user's folk theory can be used to disambiguate the indirect evidence and identify the user's folk theory with greater certainty. Once the user's folk theory has been identified, the platform designer, owner, or operator can use the folk theory to make predictions about how certain interventions will affect users' perceptions of the platform and thereby guide improvement to the platform.

Computational models of cognition provide a formal framework for modeling a user's folk theory of a sociotechnical system. Computational models of cognition have a long and rich history in psychology and cognitive science. These models have been used to study various aspects of human cognition, including perception, reasoning, decision making, and memory. In recent years, computational modeling has been increasingly used to study social phenomena. For example, computational models of social cognition have been used to study how people reason about the mental states of others, how they form and change their beliefs about the social world, and how they interact with others in social games.

In this paper, we study the problem of how a platform designer, owner, or operator can improve a digital platform by leveraging a computational cognitive model that represents users' folk theories about the platform's operation. We begin by reviewing mechanisms for content-moderation moderation on digital platforms in the context of process and outcome transparency. Next, we introduce the role that computational models of cognition can play in the design of digital platforms. We then demonstrate that the design and operation of Reddit have led to an abundance of spurious first-party suspicions of shadowbanning in cases where the mechanism was not invoked. We proceed to develop a computational cognitive model that represents users' folk theories about the antecedents and consequences of shadowbanning and predicts when users will attribute their on-platform observations to a shadowban. We then use the model to determine the interventions available to a platform designer, owner, and operator to reduce the incidence of these false suspicions and consider the implications of the approach for the design and operation of digital platforms at large.

## Shadowbanning and non-transparent content moderation

Online communities use content-moderation mechanisms to promote and enforce norms of discourse within a community and to mitigate harms that would undermine the community's purpose (Kraut & Resnick, 2012). These harms include, for example, the propagation of content that encourages and effectuates sexism, racism, radicalization, disinformation, fraud, and spam (Grimmelmann, 2015; Chandrasekharan et al., 2021). Content moderation involves screening, evaluating, categorizing, approving, promoting, removing, or hiding user-generated content based on predefined rules, guidelines, and policies (Grimmelmann, 2015). The various content-moderation mechanisms differ significantly in the transparency of their processes and outcomes (Sander, 2019; Cook et al., 2021).

Shadowbanning is a form of non-transparent content moderation where a moderator or platform owner secretly bans a user from participating in an online community or silences a user within it (Myers West, 2018; Cole, 2018). Under a shadowban, the platform is configured to filter the user's generated content (e.g., their posts and comments) so that, unbeknownst to the user, it cannot be seen by fellow community members or the public. Shadowbanning therefore produces near total non-transparency in the outcome of moderation. Shadowbanning is employed when platform owners believe that other content-moderation mechanisms with more transparent outcomes, such as suspensions or outright bans, will be evaded. For example, the user may exploit the anonymity offered by a social-media platform to create a new account and circumvent the suspension or ban (Grimmelmann, 2015). Shadowbans can have a transparent or non-transparent process, depending on whether moderators and platform owners disclose a policy that clearly articulates the conditions under which it is invoked.

## Bayesian models of cognition

Causal graphical models ("Bayes nets") are a formalism for describing the structure and strength of causal relations between events. A causal graphical model is defined by two components, a graph that describes its structure and a set of probability tables that describe the contingencies between events. The graph defining the model's structure has nodes that are random variables, with each node representing an observable or latent variable. Directed edges between the nodes represent dependencies between the events. The structure of the causal graphical model is also sometimes referred to as a directed acyclic graph, or DAG. Each node is associated with a conditional probability table, which specifies the probability of each possible outcome of the event given the outcome of the events that it depends on.

Critically, a causal graphical model can be used by a researcher in one of two ways. First, it can be used as an instrument of social science to represent the researcher's model of the world. Second, it can be used as an instrument of cognitive science to represent the researcher's model of the user's folk theories about the world. The latter formulation becomes most useful when the user's model of the world diverges from that of the experimenter, in which case it can be used to represent how users' "folk" or "intuitive" theories about the world differ from reality.

## False suspicions of shadowbanning on Reddit

We performed an empirical analysis of first-party suspicions of shadowbanning on Reddit, a social media platform organized into many diverse communities ("subreddits"). Reddit is a fitting platform for studying suspicions of shadowbanning because its administrators make extensive use of shadow content moderation, both at the user level (shadowbanning) and post level (shadow post removal). Critically, a little-known subreddit, /r/ShadowBan, enables users to directly determine whether they are shadowbanned. Though Reddit's shadowbanning mechanism operates at the level of the platform, not the subreddit, a subreddit's moderators can approve a post by an otherwise shadowbanned user so that it appears on the subreddit. The /r/ShadowBan subreddit is configured to automatically approve posts by shadowbanned users; a bot on the subreddit then replies to approved posts, informing the user whether they are shadowbanned and which of their recent posts have been removed. The present study therefore operationalizes the construct *first-party suspicions of shadowbanning* as the act of posting on /r/ShadowBan. Certainly, this definition is both too broad (because a user can post on /r/ShadowBan for any reason) and too narrow (because a user may be unaware of the subreddit or disinclined to reveal their suspicion publicly, perhaps in fear of retaliation). Even so, the subreddit provides a unique view into a phenomenon that is otherwise invisible and largely confined to private thought.

We began our analysis by empirically analyzing public logs of Reddit posts and comments (Baumgartner et al., 2020) to measure the incidence of suspicions of shadowbanning in the decade following /r/ShadowBan's creation in 2011. Over that period, 52,539 unique users registered 94,173 suspicions of shadowbanning. Of these suspicions, 3,247 (3.4%) were by shadowbanned users, whereas 90,926 (96.6%) were false suspicions by members who were not shadowbanned. Suspicious users represent a diverse sampling of Reddit users, having collectively created 5,985,042 posts and 55,976,306 comments across 152,949 subreddits. The modal subreddit that the suspicious contribute to, with over 3 million posts and comments, is /r/AskReddit, one of the largest subreddits. However, the tail is long, with 894 subreddits having at least 10,000 posts or comments by suspicious users, 4,981 subreddits having at least 1,000, and 17,322 subreddits having at least 100.

Next, we surveyed 500 Reddit users who in 2021 suspected that they were shadowbanned. To conduct the survey, we sent direct messages to Reddit users shortly after they posted to the /r/ShadowBan community, requesting that they share the basis of their suspicion. All users who posted the forum during the data-collection period were surveyed, except for those excluded according to several criteria that were applied to improve the response rate and minimize the burden of unwanted communication associated with the surveying method. First, we contacted users only if their account age was at least 6 months, which has the effect of excluding professional spammers and fraudsters, who tend to repeatedly create new accounts and respond with aggression to direct messages. Second, we contacted each user at most once and did not recontact users who checked for shadowbans multiple times. Third, we contacted only 10% of the users who were not otherwise excluded.

Our analysis of the reported bases of suspicions focused on distinguishing between process and outcome transparency (Grimmelmann, 2015) by examining perceptions of the antecedents and consequences of shadowbanning on the platform, respectively. We first report empirical results related to the *antecedents* of shadowbanning: observable factors that users interpret as causes of a moderator enacting the shadowban. The antecedent of shadowbanning most cited (by 5.5% of users) as a reason for suspicion was having written a post or comment that they believe was controversial or antagonistic. Some users (2.2%) cited the belief that moderators are strict and heavy handed in their use of shadowbans. Users also cited tens of other less frequent antecedents, often referencing specific on-platform actions by the user, fellow community members, other users, or moderators that led to conflict. Next, we report empirical results related to the *consequences* of shadowbanning: observable factors that users interpret as downstream effects of a moderator enacting the shadowban. The plurality cited cause of suspicion related to the consequences of a shadowban, reported by 16% of surveyed users, was observing less engagement with their posts and comments than expected. Alternative consequence-based routes to suspicion included observing that a post was removed (10.6%), that a comment was removed (6.2%), that multiple posts did not appear in a particular subreddit (3.1%), that their account profile was not visible to a confederate or when using a private browser session (2.7%), an inability to comment (2.7%), an inability to chat (2.2%), and a long tail of observed failures in other actions taken on the platform. (Notably, only some of the cited consequences are possible effects of shadowbans on the platform; others are technical glitches that users wrongly attributed to a shadowban.)

## Bayesian model of suspicions of shadowbanning

**Figure 1**

*A folk theory that guides users' first-party suspicions of shadowbanning.*

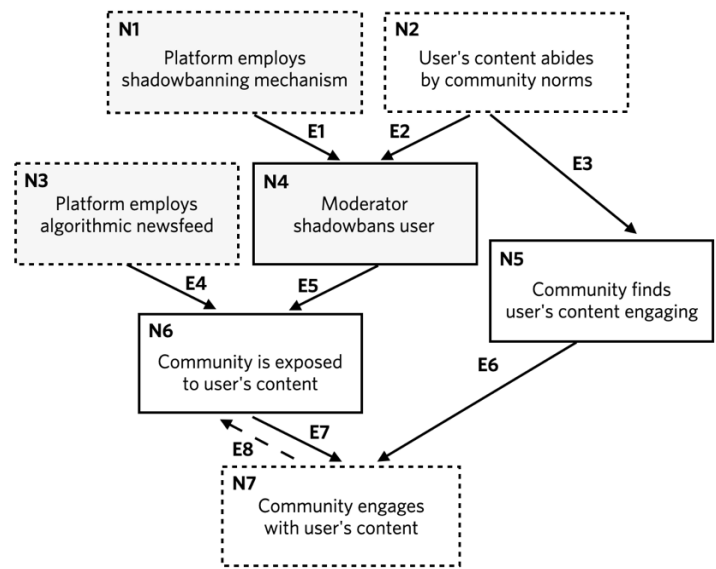

*Note.* The figure shows a diagram representing the causal model that underlies a user's first-party suspicions of shadowbanning. Nodes (N1–N7) are events and directed edges (E1–E8) are dependencies between events. Solid nodes are unobservable and dotted nodes are observable by the user. Nodes filled in grey represent events that can be intervened upon by platform designers, owners, and moderators. The edge E8 is dashed because it introduces a cycle, violating an assumption of Bayesian networks. Please note that, although event descriptions are framed actively (e.g., "Platform employs…"), nodes are events with multiple possible outcomes, including the implied negative frame (e.g., "Platform does not employ…").

In the context of shadowbanning, we might ask which events a platform designer, owner, or operator might intervene upon to affect first-party suspicions of shadowbanning. We note, however, that such an intervention would be an indirect one in that, ultimately, what is being intervened upon is only an input into the cognitive processes that construct the intuitive model under consideration here. Intervening upon

*Platform design under folk theories of sociotechnical systems*the world may or may not cause a commensurate change in the user's intuitive model of sociotechnical system. Indeed, articulation of intuitive theories is most useful as a practice when there is daylight between the ground truth of a platform's operation and the intuitive theories held by those participating in it. The intuitive theories of platform owners and operators, moderators, and users may diverge in complex ways that have ramifications for how users understand the system.

When a Bayesian network is interpreted as an intuitive theory describing the user's mental model of the world, it becomes possible for a platform owner to intervene in ways that would ordinarily be fruitless when attempting to intervene directly upon the world. In particular, the platform owner can intervene upon the user's priors over events and contingencies between events in a way that goes beyond simply setting the outcome of particular nodes. Further, the platform owner can intervene in ways that have no direct effect on the platform's operation but nonetheless has an indirect effect because of the way that they alter the inferences drawn by its users.

For an example of the alternative modes of intervention available under the intuitive-theory interpretation of a causal model, consider node N1 — whether the platform employs the shadowbanning mechanism. Under one mode of intervention, the platform owner exerts their control over the platform through a design choice about the mechanisms of moderation that are available to moderators on the platform: they can design the platform in such a way that moderators have the shadowban mechanism available to them, or alternatively, they can design the platform in such a way that no such mechanism is available. To intervene upon the user's mental model of the world, in contrast, the platform owner can publicly reveal the presence of the mechanism or attest to its absence, thereby intervening upon the user's prior over whether the platform employs the mechanism while changing nothing about the platform itself.

# REFERENCES

Baumgartner, J., Zannettou, S., Keegan, B., Squire, M., & Blackburn, J. (2020). The Pushshift Reddit dataset. In *Proceedings of the International AAAI Conference on Web and Social Media* (Vol. 14, pp. 830–839).
Brunk, J., Mattern, J., & Riehle, D. M. (2019). Effect of transparency and trust on acceptance of automatic online comment moderation systems. In *2019 IEEE 21st Conference on Business Informatics (CBI)* (Vol. 1, pp. 429-435). IEEE.
Chandrasekharan, E., Gandhi, C., Mustelier, M. W., & Gilbert, E. (2019). Crossmod: A cross-community learning-based system to assist reddit moderators. *Proceedings of the ACM on Human-Computer Interaction*, 3(CSCW), 1–30.
Cole, S. (2018, July 31). Where Did the Concept of 'Shadow Banning' Come From? *Motherboard: Tech by Vice.* https://www.vice.com/en/article/a3q744/where-did-shadow-banning-come-from-trump-republicans-shadowbanned
De Reuver, M., Sørensen, C., & Basole, R. C. (2018). The digital platform: a research agenda. Journal of Information Technology, 33(2), 124–135.
DeVito, M. A., Birnholtz, J., Hancock, J. T., French, M., & Liu, S. (2018). How people form folk theories of social media feeds and what it means for how we study self-presentation. *CHI 2018*.
Grimmelmann, J. (2015). The virtues of moderation. *Yale JL & Tech.*, 17, 42, 1–68.
Motta, M., & Callaghan, T. (2020). The pervasiveness and policy consequences of medical folk wisdom in the US. *Scientific Reports*, 10(1), 1–10.
Kraut, R. E., & Resnick, P. (2012). *Building successful online communities: Evidence-based social design*. MIT Press.
Venkatesh, V., Morris, M. G., Davis, G. B., & Davis, F. D. (2003). User acceptance of information technology: Toward a unified view. *MIS quarterly*, 425-478.*Twenty-ninth Americas Conference on Information Systems, Panama, 2023*  5